\documentclass[aps,pre,twocolumn]{revtex4}
\usepackage{graphics}
\usepackage{amsmath}
\usepackage{amssymb}
\usepackage{amsfonts}
\usepackage{graphicx}
\usepackage{color}
\usepackage{lineno,hyperref}
\begin{document}
\title{Synchronization patterns in LIF Neuron Networks: Merging Nonlocal and Diagonal Connectivity}
%\subtitle{LIF network with nonlocal and diagonal connectivity}
\author{N.~D. Tsigkri-DeSmedt} 
\affiliation{ Institute of Nanoscience and Nanotechnology, 
   National Center for Scientific Research ``Demokritos'', GR-15341 Athens, Greece }
\affiliation{Section of Solid State Physics, Department of Physics, 
National and Kapodistrian University of Athens, GR-15784 Athens, Greece}
\author{ I. Koulierakis }
\affiliation{ Institute of Nanoscience and Nanotechnology, 
   National Center for Scientific Research ``Demokritos'', GR-15341 Athens, Greece }
\affiliation{School of Electrical Engineering and Computer Science, 
National Technical University of Athens, GR-15780 Athens, Greece} 
\author{ G. Karakos}
\affiliation{ Institute of Nanoscience and Nanotechnology, 
   National Center for Scientific Research ``Demokritos'', GR-15341 Athens, Greece }
\affiliation{School of Electrical Engineering and Computer Science, 
National Technical University of Athens, GR-15780 Athens, Greece} 
  \author{ A. Provata}
\affiliation{ Institute of Nanoscience and Nanotechnology, 
   National Center for Scientific Research ``Demokritos'', GR-15341 Athens, Greece}
\date{Received: date / Revised version: date}

\begin{abstract}{{\bf Abstract:}
The effects of nonlocal and reflecting connectivities have been previously investigated in coupled 
Leaky Integrate-and-Fire (LIF) elements, which assimilate the exchange of electrical 
signals between neurons. In this work we investigate the effect of diagonal coupling inspired by
findings in brain neuron connectivity. Multi-chimera states are reported both for the simple diagonal
and combined nonlocal-diagonal connectivities and we determine the range
of optimal parameter regions where chimera states appear. Overall, the measures of
coherence indicate that as the coupling range increases (below all-to-all coupling)
the emergence of chimera states is favored
and the mean phase velocity deviations between coherent and incoherent regions become 
more prominent. A number of novel synchronization phenomena are induced as a result of the
combined connectivity. We record that for coupling strengths $\sigma < 1$
the synchronous regions have mean phase velocities lower than the asynchronous, while the opposite
holds for $\sigma > 1$. In the intermediate regime, $\sigma \sim 1$, the oscillators have common
mean phase velocity (i.e., are frequency-locked)
but different phases (i.e., they are phase-asynchronous).
Solitary states are recorded for small values of the coupling strength, which
grow into chimera states as the coupling strength increases.
We determine parameter values where the combined effects of
nonlocal and diagonal coupling generate chimera states with two different
levels of synchronous domains mediated by asynchronous regions.
%\PACS{
%      {05.45.Xt} {Synchronization; coupled oscillators}   \and
%      {87.19.lj} {Neuronal network dynamics} \and
%      {87.18.Sn} {Neural networks and synaptic communication} 
%     } % end of PACS codes
}
\end{abstract}
\maketitle
\section{Introduction}
\label{intro}
\par
%AIM OF THE CURRENT STUDY 
The neuron network of the brain is characterized by complex architecture at many scales of hierarchy, which are
still not fully recorded. The complexity of the connectivity gives rise to intricate synchronization
phenomena between different parts of the brain, which drive the exchange of information
and which are at the basis of common brain functions (cognition, memory etc.) \cite{finn:2016,logothetis:2008,poldrack:2015}. 
To unravel the influence of the neuron network connectivity in brain functionality,
large scale simulations of neuron networks are undertaken by international collaborations, 
which model the exchanges of electrical and chemical signals
in the brain \cite{vasalou:2009,li:2013}. 
The ultimate goal is to establish the missing link between brain structure 
(anatomy, neuron network connectivity, synchronization
patterns) and functionality (perception, cognition, memory) 
\cite{alonso:1989,cobb:1995,vuksanovic:2014,vuksanovic:2015,katsaloulis:2009,expert:2011,katsaloulis:2012}. 
In this direction specific models describing single neuron activity are employed, accompanied with
 connectivity rules inspired by biological 
neuron networks. Single neurons, characterized by spiking or bursting activity, are modeled by a number of 
nonlinear schemes, such as the Hodgkin-Huxley model, the Hindmarsh-Rose model, the FitzHugh-Nagumo model and the various
LIF schemes \cite{haken:2008,izhikevich:2007,wang:2018}. Commonly used linking schemes employed in the literature 
belong to two categories: deterministic or stochastic. Deterministic connectivity in neuron networks
ranges from the most local nearest-neighbor coupling, to distant 
nonlocal coupling, global coupling, field coupling schemes and even multilayer (multiplex) networks 
\cite{wang:2018,guo:2017,lv:2018,panaggio:2015,schoell:2016,majhi:2016,majhi:2017,hizanidis:2016}.  
In parallel, stochastic 
connectivity schemes have been studied, 
such as Erd{\"o}s-R{\'e}nyi networks, small world connectivity and scale free linking
\cite{schoell:2016,zhu:2014,omelchenko:2015}.
The biologically inspired models, equipped with the appropriate connectivity schemes, aim to detect the different 
levels of spatiotemporal organization and to identify synchronization patterns which are characteristic
of the functional modes of the brain. 

\par A recently discovered \cite{kuramoto:2002,abrams:2004} 
complex synchronization pattern is the chimera state; this is a collective state
characterized  by the simultaneous presence
of synchronous and asynchronous domains  
\cite{panaggio:2015,schoell:2016,rattenborg:2000,hizanidis:2014,bera:2016,andrzejak:2016,dai:2017,wu:2018}.  
Chi- mera states have been realized using a number of neuron models, such as the FitzHugh-Nagumo model 
\cite{schoell:2016,omelchenko:2015,wu:2018,omelchenko:2013,schmidt:2017,essaki:2015}, 
the Hindmarsh-Rose model of bursting neurons
\cite{majhi:2016,majhi:2017,hizanidis:2016,bera:2016,bera:2016-2,bera:2017,kundu:2018,majhi:2018},
the LIF model \cite{schmidt:2017,lucioli:2010,olmi:2010,tsigkri:2015,tsigkri:2017,kasimatis:2018} and others.
More recent studies include simulations of chime- ra states in 2D lattices 
\cite{schmidt:2017,kundu:2018}
and 3D spatial dimensions
\cite{schmidt:2017,majhi:2018,kasimatis:2018,maistrenko:2015,maistrenko:2017}. 
Another striking synchronization example is the subthreshold oscillations: the network
splits  into two coexisting domains, one domain where all
elements perform small-amplitude oscillations  at subthreshold values 
and one domain where the incoherent elements develop a
 multileveled mean phase velocity distribution \cite{tsigkri:2017}.

\par 
% PREVIOUS WORKS
In previous studies of the Leaky Integrate-and-Fire (LIF) dynamics,
a plethora of intriguing synchronization phenomena have been reported  
for coupled LIF networks with nonlocal coupling 
\cite{lucioli:2010,olmi:2010,tsigkri:2015,tsigkri:2017,tsigkri:2016}.
A few examples are briefly discussed below. 
\begin{itemize}
\item The case of negative (repulsive) coupling constant was studied using a nonlocal kernel.
This connectivity produced multichimera states
whose multiplicity (number of coherent/incoherent domains) depends on the coupling strength 
\cite{tsigkri:2015,tsigkri:2017,tsigkri:2016}.

\item In the case of positive (attracting)
nonlocal diffusive coupling, travelling waves of synchronous elements 
with constant phase difference were produced, 
while chimera states were difficult to identify. Subthreshold oscillations were observed, where the number of
oscillators staying below the threshold is statistically 
constant in time and only depends on the coupling strength \cite{tsigkri:2017}. 
Their potential values fluctuate erratically around a central point located below 
the fixed point of the uncoupled neuronal oscillator.
Note that in biological neuron networks subthreshold oscillations have been 
experimentally recorded as rhythmic fluctuations of the voltage difference between the 
interior and exterior of a neuron \cite{alonso:1989,cobb:1995}.  

\item  Hierarchical topology in the coupling was investigated and it was shown to induce
 nested chimera states and transitions between multichimera states 
with different multiplicities \cite{tsigkri:2016}.  The use of hierarchical connectivity was
motivated by previous analyses of MRI and fMRI images of the brain which reveal that the
neuron axons are distributed fractaly in the brain \cite{katsaloulis:2009,expert:2011,katsaloulis:2012}.

\item In ref. \cite{tsigkri:2017} the case of reflecting connectivity was studied, 
where each neuron is coupled with a number of other neurons located 
perpendicularly opposite, through a specific axis of the ring.
 For this connectivity, novel incoherent domains
coexisting with subthreshold elements were found.  For large values of the coupling strength
the full oscillations are restricted in
one half of the ring, while the elements of the other half perform small-amplitude,
 subthreshold oscillations \cite{tsigkri:2017}. 
 The use of the reflecting connectivity  
was inspired by the work of \cite{finn:2016}, where the recorded neuron connectivity patterns
are used as fingerprints for identifying individuals.
\end{itemize}

%The above studies revealed new complex patterns and dynamical transitions between different multichimera states 
%resulting from the combined effects of nonlinear dynamics and complex coupling.

\par
%MOTIVATION 

The results in ref. \cite{finn:2016} which motivated the use of the reflecting connectivity
in ref. \cite{tsigkri:2017} are also on the basis of the present study.
The brain neuron network analysis in \cite{finn:2016} addresses the connectivity profiles for the identification of an individual, on the basis of eight functional networks. In order to quantify edgewise contributions for the identification of an individual the authors suggested a measure termed Differential Power (DP). The DP value quantifies  how ``characteristic''
a particular edge tends to be. DP connections that characterize a)  individuals across conditions or b) across individuals  regardless of condition, appear between regions in one hemisphere and regions in the other hemisphere, 
arranged diagonally opposite to each other.  These diagonal edges (connections) are characteristic of each individual
and are good candidates for identifying individuals (fingerprints). In the opposite case, the measure
Group Consistency (denoted by $\Phi$) 
was introduced which quantifies the consistency of a connection within a subject and across a group. 
 $\Phi$ connections that are consistent across a group, appear to link the two hemispheres perpendicularly to the plane separating them. They are common to all individuals, they constitute
the major parts of connections and consequently cannot be used to identify individuals.

\par 
% PRESENT OUR PRESENT WORK
Following the study on the reflecting connectivity   which gave rise to subthreshold oscillations \cite{tsigkri:2017}
and which was inspired by the $\Phi$ connections of Ref. \cite{finn:2016}, 
we now turn to diagonal and nonlocal coupling schemes, inspired by the DP connections in the same reference.
 Each neuron will be now
connected with $2R_{diag}$ neurons symmetrically distributed across a standard mirror diameter of the ring
(symmetry about the center). We also consider
combined nonlocal-diagonal linking were each element is linked with a number of neurons $R_{nl}$ on 
its left and $R_{nl}$ on its right in addition to the diagonal ones. 
This is inspired by the fact that neurons can be simultaneously interconnected in their structural areas 
and functionally linked 
with other regions of the brain as indicated in \cite{omelchenko:2011}. 
 The novelty of the present study lies not only 
 in the use of the particular, biologically-inspired connectivities, but also on the novel synchronization
phenomena induced. Namely, a) the appearance of solitary states which gradually give rise to chimera states as the
coupling strength increases, b) the interesting bi-leveled chimeras where two levels of synchronization appear and the incoherent
regions are located between them and c) the inversion of coherence for large values of the coupling strength $\sigma$: 
for small (large) values of $\sigma$ the frequency of the coherent regions is lower (higher) than the incoherent ones.

\par 
% WORK ORGANIZATION
This work is organized as follows: 
In section \ref{sec:LIF} we recapitulate the main  properties of  
the single LIF model and we introduce the nonlocal and diagonal coupling.
In section \ref{sec:diagonal} we explore the network synchronization motifs when simple diagonal connectivity is assigned.
In section \ref{sec:combined} A and B we scan the parameter space and discuss the effects 
of the combined connectivity scheme on the dynamical behavior of the system.
In section \ref{sec:critical} we scan  the parameter region where bi-leveled synchronous domains appear
and we discuss the mechanisms responsible for this effect.
In the concluding section we propose future steps of this study and we summarize our main results.

\section{Leaky Integrate-and-Fire Model with linear coupling}
\label{sec:LIF}

To study the dynamical evolution and the synchronization phenomena in a system
of $N$ neurons we use the Leaky Integrate-and-Fire (LIF) model 
\cite{lapicque:1907a,lapicque:1907b}, which describes the dynamics of single neurons. 
The LIF model exhibits an exponential increase of the potential followed by  an abrupt resetting
to the rest state.
The evolution of the membrane potential $u(t)$ of a single neuron is represented by the following equations: 

\begin{subequations}
\begin{equation}
\label{eq1a} 
\frac{du(t)}{dt}=\mu-u(t)
\end{equation}
\begin{equation}
\lim_{\epsilon \to 0^{+}}u(t+\epsilon ) \to u_0, \>\>\> {\rm when} \>\> u(t) \ge u_{\rm th}
\label{eq1b}
\end{equation}
\label{eq01}
\end{subequations}

\noindent where $\mu$ defines the spiking rate and $u_{\rm th}$ denotes the threshold value of the potential, which
when exceeded, $u$ is restored to its ground state $u_0$.
Assuming, without loss of generality, that at time $t=0$ the potential is at $u_0$ and that the resetting to
the ground state is abrupt, the  period $T_s$ of the single neuron is calculated from Eq.~\ref{eq01} as 
$T_s=\ln \left[ (\mu -u_0)/(\mu - u_{\rm th} ) \right] $.

\par
\begin{figure}[h!]
\includegraphics[clip,width=0.9\linewidth,angle=0]{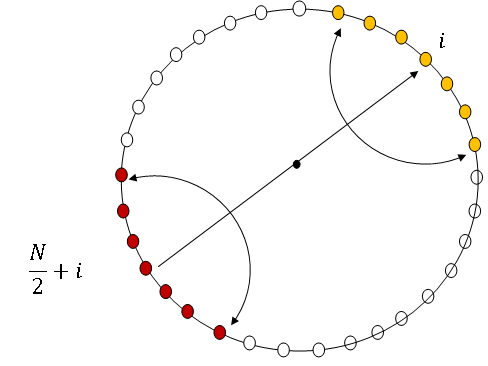}
\caption{\label{fig-diagnl} (Color online)
Nonlocal and diagonal connectivity scheme. In the diagonal connectivity scheme
every element $i$ is linked with $2R_{\rm diag}+1$ neighbors across the diagonal of the ring
(red, dark nodes),
while in the combined connectivity scheme $i$ is linked, additionally,
with $R_{\rm nl}$ other elements to its left and $R_{nl}$ elements to its right (yellow, grey nodes). 
 The connectivity range
here is $R=3$ (with $R_{\rm diag}=R_{\rm nl}$). }
\end{figure}

 For the network of interacting neurons we use linear coupling between neurons $j$ and $i$
and the coupling is defined by the adjacency or coupling matrix $\sigma(i,j)$. 
 For a number of $N$ LIF elements, each of them having potential $u_i(t)$, $i=1, \cdots N$ and
connected on a ring topology the network dynamics is described by the following equations:

\begin{subequations}
\begin{equation}
\label{eq6a} 
\frac{du_i(t)}{dt}=\mu-u_i(t)-\frac{1}{N_i} \sum_{j=1}^{N_i} \sigma (i,j)
\left[ u_j(t)-u_i(t)\right] 
\end{equation}
\begin{equation}
\lim_{\epsilon \to 0^{+}}u_i(t+\epsilon ) \to u_0, \>\>\> {\rm when} \>\> u_i(t) \ge u_{\rm th},
\label{eq6b}
\end{equation}
\label{eq06}
\end{subequations}

\noindent where $N_i \le N$ is the total number of elements that are linked to element $i$. 
The parameters $N_i$ may be different for each element but in this study
 we assume that $N_i=N_c$ is kept constant for all elements.
Moreover, in Eq.~(\ref{eq06})  we assume that all elements have as common parameters (properties) the threshold 
potential $u_{\rm th}$, the spiking rate $\mu$, the rest potential $u_0=0$,
while they start from different (usually random) initial conditions (potentials). Regarding the
connectivity matrix  $\sigma(i,j)$
we consider here the following two cases: simple
diagonal connectivity and combined nonlocal-diagonal connectivity. In the first case 
 the matrix element $\sigma_d (i,j)$ linking node $j$ to $i$ takes the form:
\small
\begin{eqnarray}
\sigma_d(i,j)=
\left\{
  \begin{aligned}
    \sigma &\>\> {\rm for} \> (\frac{N}{2}+i-R_{\rm diag}) \le j\le (\frac{N}{2}+i+R_{\rm diag}) \\
    0 &\>\> {\rm otherwise}
  \end{aligned}
\right.
\label{eq02}
\end{eqnarray}
\normalsize
\noindent where $\sigma$ is a positive constant and 
all indices are understood $\mod (N)$. In Fig.~\ref{fig-diagnl} every element  is
only linked to $2R_{\rm diag}+1$ neighbors (red, dark-colored) across the diagonal of the ring.
The ratio of the linked
elements to the total system size is the coupling ratio which is denoted by $d=(2R_{\rm diag}+1)/N$ 
and will be used as one of the system
parameters in the sequel.

\par In the case of the combined (nonlocal-diagonal) connectivity the adjacency matrix $\sigma_c(i,j)$ takes the form
\small
\begin{eqnarray}
\sigma_c (i,j)=  
\left\{
  \begin{aligned}
    \sigma &\>\> {\rm for} \>\> (\frac{N}{2}+i-R_{\rm diag}) \le j \le (\frac{N}{2}+i+R_{\rm diag})  \\
     \sigma &\>\> {\rm for} \>\> i-R_{\rm nl} \le j\le i+R_{\rm nl}  \\
    0 &\>\> {\rm otherwise.}
  \end{aligned}
\right.
\label{eq03}
\end{eqnarray}
\normalsize
Here, similarly to Eq.~\ref{eq02}, $\sigma$ is positive and 
all indices are understood $\mod (N)$.  In this scheme every element is 
linked to its $R_{\rm nl}$ neighbors on its left, to its $R_{\rm nl}$ neighbors on its right and also
to $2R_{\rm diag}+1$ neighbors across the diagonal of the ring, i.e., all red (dark) and yellow (grey) colored
elements in Fig.~\ref{fig-diagnl}. This
means that every element will be linked to a total of $(2R_{\rm diag}+2R_{\rm nl}+1)$ nodes. To avoid introducing
many parameters we hereafter
consider the case where $R_{\rm diag}$=$R_{\rm nl}=R$, 
thus the number of total elements to which each element is linked, is equal to $4R+1$. The 
 coupling ratio in the case of combined connectivity is $d=(4R+1)/N$.

\par Synchronization properties are quantitatively described by the mean phase velocity $\omega_i$
 of element $i$. If the system is integrated for time $\rm{\Delta T} $, we compute the number of full
cycles $k_i$
which element $i$ has completed during time ${\rm \Delta T}$.
 The mean phase velocity, or average frequency $ \omega_i$ of element $i$ 
is then estimated as follows \cite{omelchenko:2013}: 
\begin{eqnarray}
\omega_i=\frac{2 \> \pi \> k_i}{\rm{\Delta T}}
\label{eq07}
\end{eqnarray}
\noindent Qualitatively, synchronization patterns and chimera states are represented by the space-time plots.
Space-time plots are color coded plots that show the evolution of the oscillator potentials in time and in space. 
These are particularly useful for monitoring transitions between different
synchronization patterns or travelling chimeras, when the mean phase velocities do not give meaningful results.

\par We complement our study with the calculation of the relative size of the incoherent parts $N_{\rm incoh}$ and with the
extensive cumulative size $M_{\rm incoh}$ which represents the degree of incoherence of the chimera states \cite{omelchenko:2015}.  The quantities  $N_{\rm incoh}$ and $M_{\rm incoh}$ are defined as follows:

\begin{subequations}
\begin{equation}
\label{eq8a}
N_{\rm incoh}=\frac{1}{N} \sum_{i=1}^{N} \Theta (A)
\end{equation}
\begin{equation}
A=
\left\{
  \begin{aligned}
\omega_i-\omega_{\rm coh}-c,\> {\rm when} \>\>\omega_{\rm coh}<\omega_{\rm incoh}\\
\omega_{\rm coh}-\omega_i-c,\> {\rm when} \>\>\omega_{\rm coh}>\omega_{\rm incoh}
\label{eq8b}
\end{aligned}
\right.
\end{equation}
\label{eq08}
\end{subequations}

\begin{eqnarray}
M_{\rm incoh}= \sum_{i=1}^{N} \mid  (\omega_i-\omega_{\rm coh}) \mid .
\label{eq09}
\end{eqnarray}

\noindent In Eqs.~\ref{eq08} and ~\ref{eq09}, $\omega_i$ denotes the mean phase velocity of the $i-$th element and
$ \omega_{\rm coh}$ the common mean phase velocity of the coherent parts. 
$\Theta$ is the step function which takes the value 1 
when its argument is positive 
and zero otherwise. To account for the fluctuations at the level of coherent parts
 a small tolerance is included in the calculations when computing the measures $M_{\rm incoh}$ and $N_{\rm incoh}$. 
This is represented by parameter $c$ which is set to 0.05. Note that the case
of complete synchronization is obtained when $M_{\rm incoh}=0$ and $N_{\rm incoh}=0$.

\begin{figure}[h!]
\includegraphics[clip,width=1.1\linewidth,angle=0]{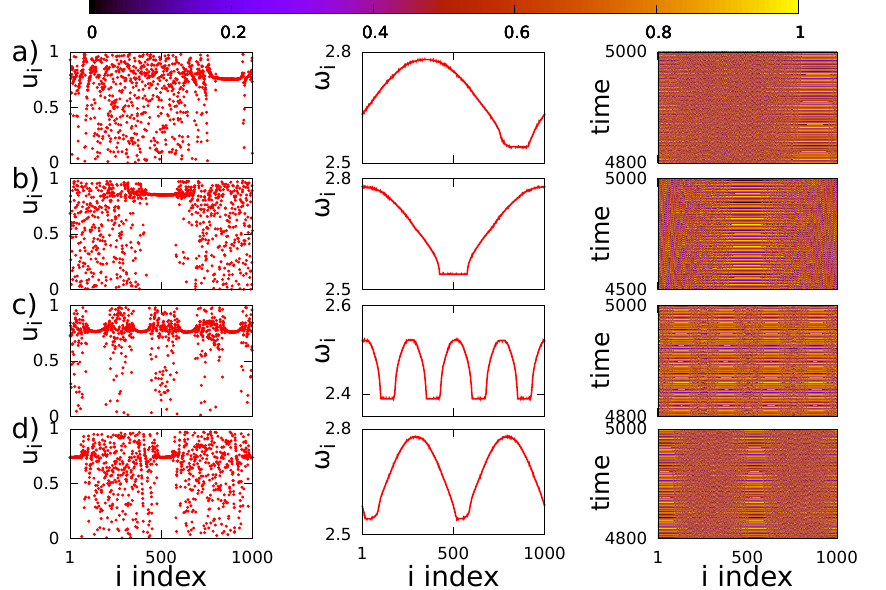}
\caption{\label{fig2} (Color online)
LIF system with different connectivity schemes and for the same number of connection, $R_{total}=600$ links
($d=0.6$):
Typical snapshots (left column), mean-phase velocity (middle column) and space-time plots (right
column). a) Nonlocal connectivity, b) reflecting connectivity, c) diagonal connectivity and d) combined nonlocal-diagonal connectivity.
Other parameters are common: $\sigma=0.6$,  $N = 1000$, $\mu = 1$
and $u_{\rm th} = 0.98$. All realizations start from the same initial
conditions, randomly chosen between 0 and $ u_{\rm th}$.
}
\end{figure}

\par  Previous studies have shown that different connectivity schemes result in different dynamical behavior. 
We present, in Fig.~\ref{fig2},
typical results for different connectivity schemes, namely
 for a) nonlocal connectivity, b) reflecting connectivity c) diagonal connectivity and d) 
combined nonlocal and diagonal connectivity. Note that all four cases are simulated under common parameters and only
the links are placed differently in the network. This figure demonstrates explicitly how
influential the connectivity in synchronization is.
In the next sections our focus will be on how simple diagonal and combined
nonlocal-diagonal connectivities affect the dynamics of the LIF network. 
Throughout this work the parameter space of the coupling range and the coupling strength is scanned. Other working parameters
are kept to values $\mu =1.0$, $u_{\rm th}=0.98$, $u_0=0.0$, common for all elements. 
We assume that the elements always start with initial potentials randomly distributed in the interval $\left[ 0, u_{\rm th}\right]$. 

\section{Effects of diagonal coupling}
\label{sec:diagonal}
Although simple diagonal coupling is not common in natural networks where the elements tend to connect to their neighbors
before linking with distant nodes, for reasons of completeness we choose to study first the case of simple diagonal
connectivity before investigating the combined one. As in the case of usual, nonlocal connectivity we keep symmetry in the
linking arrangement.  In a ring network of size $N$, the elements coupled to a node $i$ are symmetrically placed around the
diagonal node $N/2+i$ (modulo $N$), see Fig.~\ref{fig-diagnl} (red, dark nodes). The coupling matrix is given by
Eq.~\ref{eq02}, with $R_{\rm diag}=R$ common to all elements. In the next two subsections we present 
the synchronization phenomena which are induced as we vary the coupling strength,
sec.~\ref{sec:diagonal-strength} and the coupling range, sec.~\ref{sec:diagonal-range}.

\subsection{Diagonal connectivity with coupling strength variations}
\label{sec:diagonal-strength}
\begin{figure}[h!]
\includegraphics[clip,width=1.05\linewidth,angle=0]{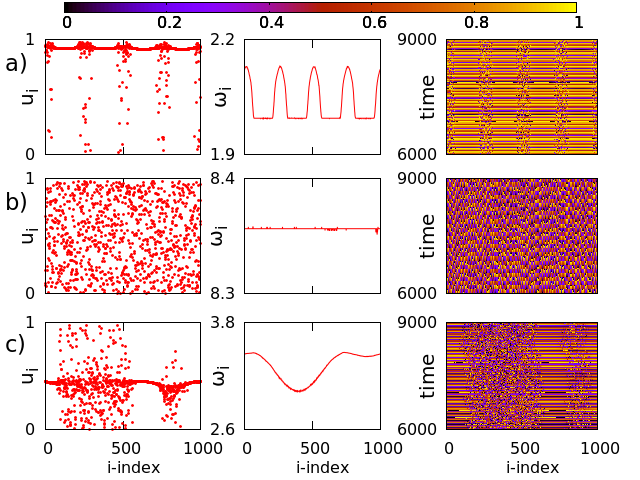}
\caption{\label{fig:d3} (Color online)
LIF system with simple diagonal connectivity:
Typical snapshots (left column), mean-phase velocity (middle column) and space-time plots (right
column). a) $\sigma=0.4$, b) $\sigma=1.0$, c) $\sigma=1.4$.
Other parameters are  R=300 (d=0.601), N = 1000, $\mu = 1$
and $u_{\rm th} = 0.98$. All realizations start from the same initial
conditions, randomly chosen between 0 and $u_{\rm th}$.
}
\end{figure}

F or diagonal connectivity, multi-chimera states are observed whose multiplicity depends on the size of the 
coupling strength. As an example, for fixed coupling range $R=300$ we plot in Fig.~\ref{fig:d3} the potential profiles (left),
the mean phase velocities (middle) and the spacetime plots (right), for three different values of $\sigma =0.4,\> 1,\> 1.4$.
We recognize a four-headed chimera for $\sigma =0.4$, an asynchronous state for $\sigma =1.0$ and an
asymmetric chimera for $\sigma =1.4$.
It is remarkable that for $\sigma < 1 $ the synchronous regions have lower mean phase velocities than the asynchronous ones,
while the opposite is true for $\sigma > 1 $. In the case $\sigma = 1$  all oscillators have the same mean phase velocity,
but different phases; they are frequency-synchronous but remain phase-asynchronous. The reversion of the synchronous-asynchronous
levels has been also observed by Omelchenko et al. \cite{omelchenko:2015-2} for nonlocally coupled the Van der oscillators,
when the coupling constant exceeds a certain threshold. The present results show that  it
is a general property of the LIF model and has been previously reported in Ref.~\cite{schmidt:2017} for the 2D connectivity.
The same effect also holds
for the combined (nonlocal-diagonal) connectivity, as we will see in sec.~\ref{sec:combined}.

\par
\begin{figure}[h!]
\includegraphics[clip,width=1.0\linewidth,angle=0]{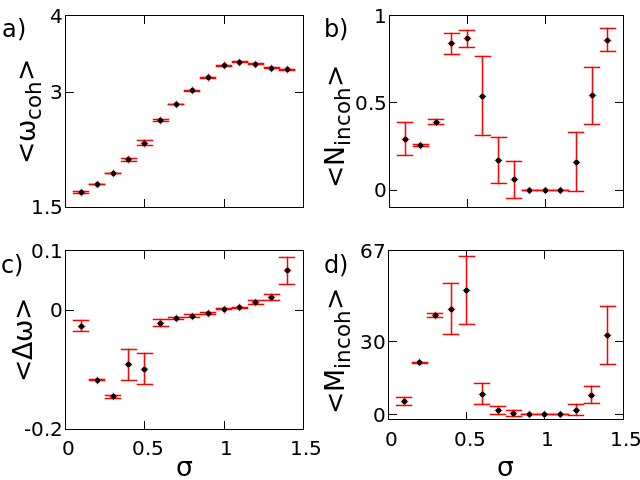}
\caption{\label{fig:d4} (Color online)
LIF system with simple diagonal connectivity:
Measures of coherence and incoherence for different values of the coupling strength and for $R=200$.
Other parameters as in Fig.~\ref{fig:d3}. 
}
\end{figure}

 The reversion of the synchronization levels is clearly shown in Fig.~\ref{fig:d4}, where all measures of coherence
present transition points around $\sigma \sim 1$. In panel ~\ref{fig:d4}a
the average $<\omega_{\rm coh} >$ of the coherent regions increases as a function of $\sigma$ for $\sigma <1$.
 In the vicinity
of $\sigma \sim 1$ an extremum is observed, while for $\sigma >1$ the average $<\omega_{\rm coh} >$ slightly decreases.
Similarly,  in Fig.~\ref{fig:d4}c the difference $\Delta\omega $ between coherent and incoherent regions
cross the 0-axis for $\sigma \sim 1$. Around the same transition point the number of incoherent elements $< N_{\rm incoh}>$
vanishes (see Fig.~\ref{fig:d4}b) and so does the index $< M_{\rm incoh}>$ which accounts for the area 
(in the $\sigma - \omega $ graph) between coherent and incoherent regions (see Fig.~\ref{fig:d4}d).

\subsection{Diagonal connectivity with coupling range variations}
\label{sec:diagonal-range}

\begin{figure}[h!]
\includegraphics[clip,width=1.05\linewidth,angle=0]{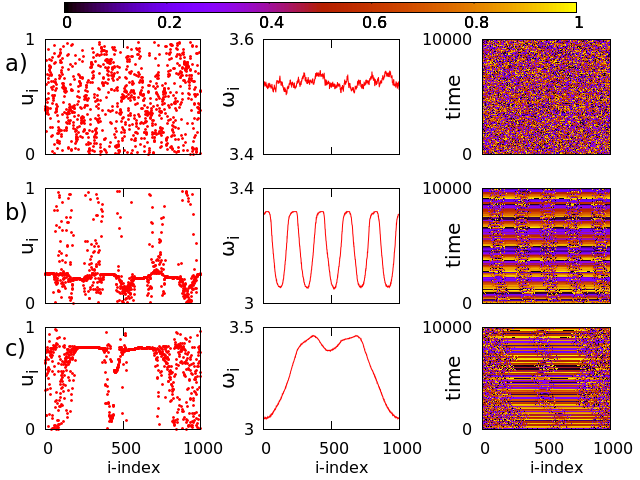}
\caption{\label{fig:d5} (Color online)
LIF system with simple diagonal connectivity:
Typical snapshots (left column), mean-phase velocity (middle column) and space-time plots (right
column). a) R = 10 (d=0.041), b) R = 250 (d=0.501), c) R = 330 (d=0.661).
Parameters are $\sigma =1.4$ and others as in Fig.~\ref{fig:d3}.
}
\end{figure}

As previously shown in the literature the coupling range plays a special role in controlling the
multiplicity of the chimera state \cite{omelchenko:2013,tsigkri:2016,hizanidis:2015}. In particular,
as we increase the coupling range the number of coherent(incoherent) regions decreases. The same is
true here, although the position of the coupled nodes is diagonally opposite to one another.
In fact, for small values of $R$ no chimera states are observed, see Fig.~\ref{fig:d5}a. 
The same is true in the case of
classical nonlocal connectivity, when the elements are linked
only with a few nearest neighbors in their immediate vicinity
\cite{omelchenko:2015,omelchenko:2013,tsigkri:2015,tsigkri:2016}. As the value of $R$ increases
the coupling causes organization in small regions and 
a number of coherent/incoherent regions are formed, see Fig.~\ref{fig:d5}b. As the coupling range
increases further, the multiplicity of the chimeras decreases as can be seen in Fig.~\ref{fig:d5}c
(see also Fig.~\ref{fig:d3}c for comparison). The appearance of incoherent regions of different
sizes is a new effect, which could be attributed
 due to the delocalization of the interactions (diagonal linking) and to an instability created
in the middle of a coherent region giving rise to ``rebels'' forming gradually incoherent domains.
 
\begin{figure}[h!]
\includegraphics[clip,width=1.0\linewidth,angle=0]{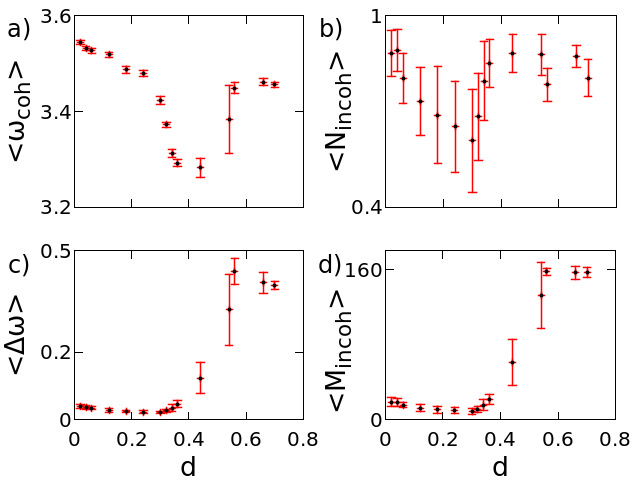}
\caption{\label{fig:d6} (Color online)
LIF system with simple diagonal connectivity: 
Measures of coherence-incoherence for different values of the coupling range. 
$d$ is the coupling ratio, $\sigma=1.4$ and
all other parameters are as in Fig.~\ref{fig:d3}.
}
\end{figure}

The measures of coherence also indicate here transitive behavior in the parameter region $150 <R < 300$.
In particular, for small values of the coupling range $R$, all elements present common $\omega$'s with
some fluctuations
as can be seen from Fig.~\ref{fig:d6}c, indicating that $\Delta \omega \simeq 0$. Regardless of the
common frequency, in this
parameter range the oscillator phases remain asynchronous 
as can be seen in Fig.~\ref{fig:d5}a (left and right panels).
As the coupling range increases
the average mean phase velocity decreases due to the effort of the elements to organize, see  Fig.~\ref{fig:d6}a.
Around $R\sim 150$ a transition region begins, where all measures increase, while
after $R\sim 300$ all measures do not show significant change with $R$.

\section{Effects of combined nonlocal and diagonal coupling}
\label{sec:combined}

\par Turning to the case of combined nonlocal-diagonal connectivity which is more relevant in
biological processes, in the next subsections 
we present the synchronization phenomena
observed as we vary the coupling range (sec.~\ref{sec:radius}) and the coupling strength (sec.~\ref{sec:sigma}).
In particular, parameter changes induce solitary states and a variety of chimeras 
which differ mainly in their multiplicity and the size of
the (in)coherent domains.

\begin{figure}[h!]
\includegraphics[clip,width=1.2\linewidth,height=1.0\linewidth,angle=0]{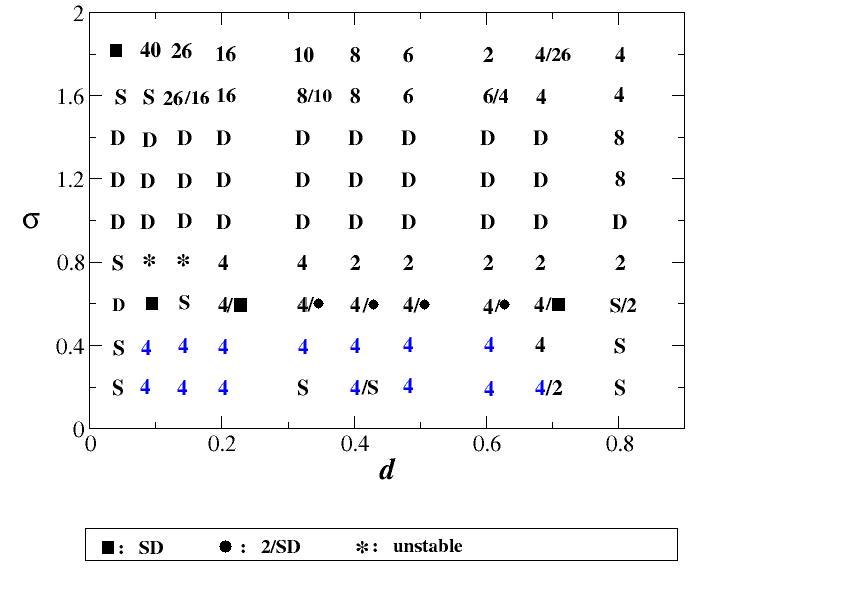}
\caption{\label{fig-map}
Mapping of the chimera multiplicity on the $(d-\sigma)$ parameter space. The numbers correspond to
the multiplicity of the chimeras, the letters ``D" and ``S" denote desynchronization and  synchronization
respectively, and the stars mark unstable chimera states. The blue color denotes solitary states;
these states are found for $\sigma \le 0.4$, for most values of $R$ and have multiplicity 4.
For other notation see the legend.}
\end{figure}

 In Fig.~\ref{fig-map} we present collectively in the ($d-\sigma$) parameter space the quantitative results
 of our study, while the details 
of the simulations and discussion of the results are given in secs.~\ref{sec:radius} and ~\ref{sec:sigma}. 
For different pairs
of ($d$, $\sigma$) on the map we indicate the multiplicity of the corresponding chimera. 
For example, in position ($d=0.2, \> \sigma=0.8$) 
a four-headed chimera is observed and thus the number ``4'' appears in the corresponding position of the map. 
In the cases of smaller
$\sigma$ , i. e. $\sigma=0.2-0.4$ the patterns are chimera-like; they consist of mostly synchronous elements
with spurious isolated
unsynchronized elements (or small asynchronous regions) also known as ``solitary states''
\cite{kapitaniak:2014,premalatha:2016,jaros:2018}. In Fig.~\ref{fig-map} solitary states are denoted with blue color;
they are found for $\sigma \le 0.4$ and their multiplicity is 4. 
In some cases two numbers appear in the same position of the plot; this means that different 
simulations resulted in different states: 
see for example ($d=0.6,\> \sigma=1.6$), where the indication ``6/4'' appears in the 
corresponding position. In this case some simulations resulted 
in two-headed chimeras while others in four-headed ones. Note that these 
double values appear in the boundaries between domains where multiplicity changes.
 The character ``S" corresponds to complete synchronization, the character ``D" to 
complete desynchronization, while the stars, $*$, denote unstable chimeras
 (meaning chimera are formed but do not settle) for these simulation times. 
All results in Fig.~\ref{fig-map} are based 
on 14 independent simulations for each pair ($d,\sigma$).
More detailed 
description and quantitative features of the patterns follow in the next two subsections.

\subsection{Combined Connectivity with variation of the coupling range}
\label{sec:radius}

\par
In this subsection we discuss the results on the morphology of chimera states and other
synchronization patterns by systematically changing the coupling range $R$.

\begin{figure}[h!]
\includegraphics[clip,width=1.1\linewidth,angle=0]{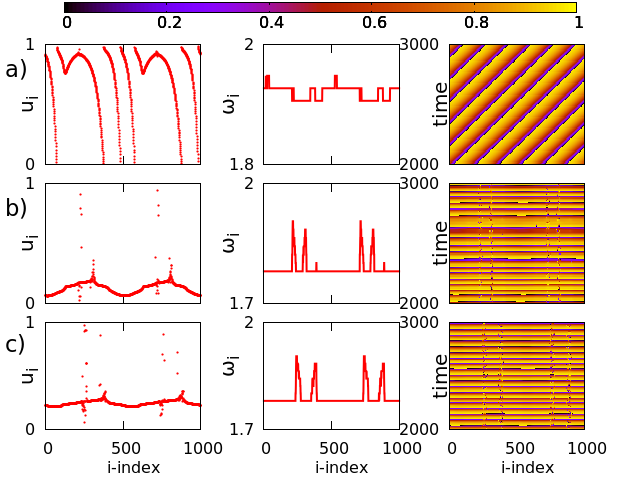}
\caption{\label{fig-R} (Color online)
LIF system with superposition of nonlocal  and diagonal connectivity:
Typical snapshots (left column), mean-phase velocity (middle column) and space-time plots (right
column). a) $R = 10 \quad(d=0.041)$, b) $R = 100\quad (d=0.401)$ and c) $R = 170 \quad (d=0.681)$.
Other parameters are $\sigma = 0.4$, N = 1000, $\mu = 1$.
and $u_{\rm th}$ = 0.98. All realizations start from the same initial
conditions, randomly chosen between 0 and $u_{\rm th}$.  }
\end{figure}
\par
For small values of  $R < 20$ $(d < 0.08)$ and $\sigma \le 0.4$, complete $\omega$-synchronization
settles in the system (see Fig.~\ref{fig-R}a), as in the case of simple diagonal coupling.
 By gradually increasing the values of the coupling range,
we first observe a chimera-like state, meaning the simultaneous existence 
of a dominant synchronization and a few asynchronous ``rebels'', see Fig.~\ref{fig-R}b . 
As we will see in the next section, sec.~\ref{sec:sigma}, keeping $R=$const.
and gradually increasing the $\sigma$ values,
these rebels turn into bigger incoherent clusters,
giving rise to typical chimera states.
Rebels, also known as solitary states have been previously reported in \cite{jaros:2018} for the Kuramoto model with inertia
and for small values of the coupling range (see also \cite{kapitaniak:2014,premalatha:2016,gopal:2018}). 
In the LIF model, as $R$ increases beyond $R=150\quad (d=0.6)$, the solitaries grow in
numbers forming aggregates, and ultimately turn into the incoherent domains of the classical chimeras.
Comparing Figs.~\ref{fig-R}b and c, we observe how the mean phase velocity profile of the incoherent regions broadens
as $R$ increases. 
For even larger values of the coupling range,
e.g. $R=200$, chimera states become more prominent or the system returns to a state of complete 
synchronization.
\par
The spatiotemporal evolution of the system for a small value of the coupling strength $\sigma=0.4$
and for a variation of the coupling radius $R$ in the nonlocal and in the diagonal neighborhoods, are demonstrated in Fig.~\ref{fig-R}. As previously, for small values of $R$ we
observe complete synchronization while for intermediate $R$ values  a few
desynchronized solitaries appear. These solitaries are organized in four different groups
and  can be spotted in the spacetime plots (right panels) and the mean phase velocity
plots (middle panels). The spacetime plots are depicted for a small time window for image resolution purposes. Looking closer
to the mean phase velocity profiles  we distinguish that in the region where the solitaries appear
 the $\omega$'s take values in 
the range $\omega \in (1.8-1.9)$.

\begin{figure}[h!]
\includegraphics[clip,width=1.2\linewidth,angle=0]{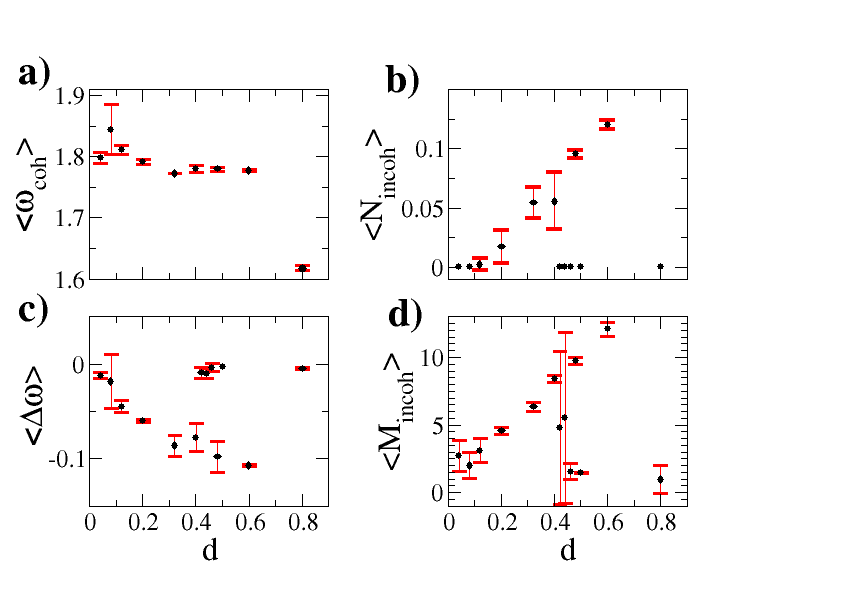}
\caption{\label{fig-coh2} (Color online)
Measures of coherence and incoherence for different values of the coupling ratio $d$. $\sigma=0.4$ and
all other parameters are as in Fig.~\ref{fig-R}.   
}
\end{figure}

\par Figure ~\ref{fig-coh2}  displays measures of coherence and incoherence as a function of the
coupling range, for small $\sigma$ values $(\sigma =0.4)$. 
Each point represents an average over
fourteen different initial conditions. 
Figure ~\ref{fig-coh2}c demonstrates that synchronization settles in the system for very small 
and for very large values of the coupling range $R$. The synchronization at large coupling ranges is not
surprising, since the system reaches an all-to-all connectivity in this limit.
In the region of small $R$ the mean phase velocity of the coherent regions is found approximately constant
(around $\omega \sim 1.8$), while it drops as we approach
the all-to-all connectivity (see Fig. ~\ref{fig-coh2}a). 
As $(4R+1) \to N$ the system slows down while synchronizing and this is expressed 
by a smaller value of the mean phase velocity (e.g.  for
$d=0.801$ where $\omega \sim 1.6$). Figure ~\ref{fig-coh2}b shows that the population of
the incoherent elements increases as the coupling ratio $d$
increases, up to $d=0.801$ when it drops back to zero, which reflects the complete synchronization
in the large $d$ limit.
 The difference of the
$\omega$'s of the coherent and incoherent elements, $\Delta\omega$, increases gradually as the 
coupling ratio $d$ increases.
Since the $<\omega_{\rm coh}>$ remains approximately constant with the increase of d, 
we conclude that the incoherent domains
become more active. As expected, $\Delta\omega$  is zero for complete synchronization (i.e. $d=0.041,\>\> d=0.801$). This
is also reflected by the evolution of $<N_{\rm incoh}>$ and $<M_{\rm incoh}>$. Both $<N_{\rm incoh}>$ and
 $ <M_{\rm incoh}>$ receive their maximum values
when  $\Delta\omega$ is also maximum, thus suggesting that the incoherent domains become more pronounced. Overall,
 the $<N_{\rm incoh}>$ and
 $ <M_{\rm incoh}>$ absolute values are low indicating that the small $\sigma$ values form ``weak-chimeras'', 
consisting mainly of synchronous regions interrupted by few``rebels''.

\begin{figure}[h!]
\includegraphics[clip,width=1.1\linewidth,angle=0]{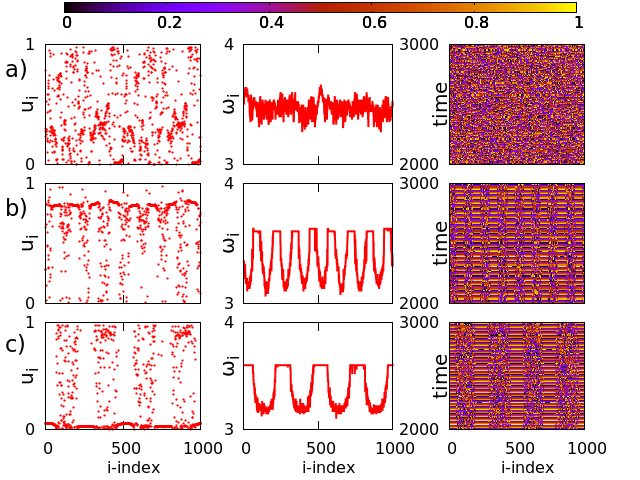}
\caption{\label{fig-R2} (Color online)
LIF system with superposition of nonlocal  and diagonal connectivity:
Typical snapshots (left column), mean-phase velocity (middle column) and space-time plots (right
column). a) R = 10 (d=0.041), b) R = 100 (d=0.401) and c) R = 150 (d=0.601).
Other parameters are $\sigma = 1.6$, $N = 1000$, $ \mu = 1$
and $u_{\rm th}$ = 0.98. All realizations start from the same initial
conditions, randomly chosen between 0 and $u_{\rm th}$. 
}
\end{figure}

 Let us now discuss the spatiotemporal evolution of the system for a greater value of the coupling strength, $\sigma=1.6$, demonstrated in Fig.~\ref{fig-R2}. 
Again for very small values of the coupling range $R$ the system does not support chimera states, see Fig.~\ref{fig-R2}a. As in 
the case of Fig.~\ref{fig-R}a, when the coupling strength
is weak and the coupling range is very small the coupled elements do not  significantly alter the local dynamics;
instead, they all
keep common $\omega $'s very close to the uncoupled system while their phases remain asynchronous.
 As the coupling strength
increases the coupled elements have a stronger bond and the system is driven into forming
chimeras. Furthermore, we note that as the coupling range increases the multiplicity of
the chimera states decreases. For example, for $R=100$ we observe an eight-headed chimera (Fig.~\ref{fig-R2}b)
 and for $R=150$ we
observe a four-headed chimera (Fig.~\ref{fig-R2}c). 
\par
Returning to the map, Fig.~\ref{fig-map}, different $(d-\sigma )$  pairs result in different multi-headed chimeras.  
We note that some pairs serve as borderlines between the different multi-head chimera areas or pure synchronous 
and asynchronous regimes. 
The steady state of these ``borderline" pairs may be affected by the states of surrounding regions and usually
result into one or another multi-headed chimera, depending on the initial
conditions. 
An example of this behavior is reported in the case of Fig.~\ref{fig-R2}c, 
in which according to Fig.~\ref{fig-map} the system sets
 to either a six-headed or a four-headed chimera (shown in this figure).

\begin{figure}[h!]
\includegraphics[clip,width=1.2\linewidth,angle=0]{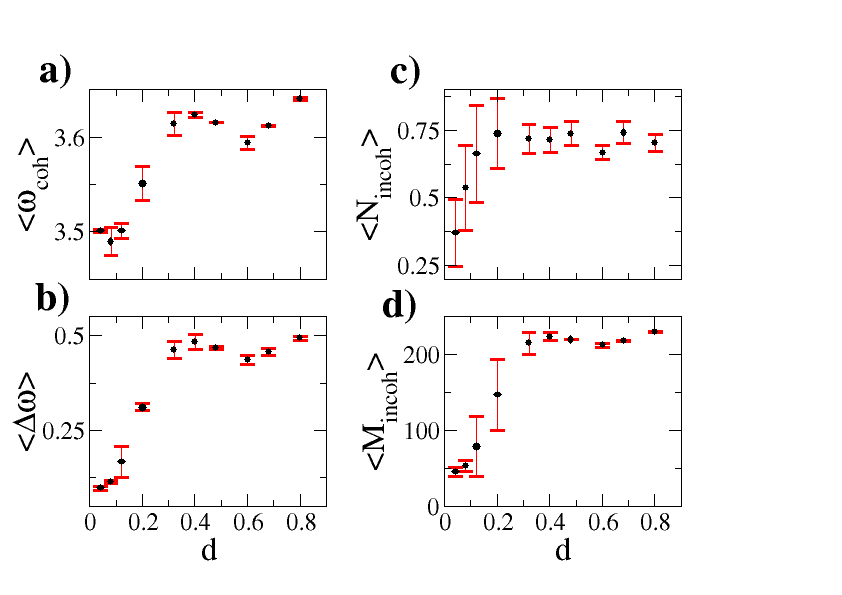}
\caption{\label{fig-coh3} (Color online)
Measures of coherence and incoherence for different values of the coupling ratio $d$. $\sigma=1.6$ and
all other parameters are as in Fig.~\ref{fig-R}. Averages are taken over 14 initial conditions. 
}
\end{figure}

\par The measures of coherence and incoherence in this case are demonstrated in Fig.~\ref{fig-coh3}. Overall, 
we note that the
values of the coherent mean phase velocities are higher compared to the ones in Fig.~\ref{fig-coh2}. This along with the plots in 
Fig.~\ref{fig-R2}, shows that for higher values of the coupling strength the coherent regimes have higher values of the
mean phase velocity. When $\langle \omega_{\rm coh} \rangle$ receives its maximum and 
minimum value the same happens accordingly
for the $\Delta\omega$. $<N_{\rm incoh}>$ and $<M_{\rm incoh}>$ receive higher values 
 than in Fig.~\ref{fig-coh2}. This is expected as the
incoherent clusters are bigger in size and chimera states are formed. Furthermore, $<N_{\rm incoh}>$ and $<M_{\rm incoh}>$ 
increase
as the coupling ratio $d$ increases up to $d=0.401$ and thenceforth remain approximately constant. This means that
the incoherent population does not significantly change after a certain point.

\par Comparing Figs.~\ref{fig-coh2} and ~\ref{fig-coh3},  
it becomes evident that for smaller values of the coupling strength (Fig.~\ref{fig-coh2}) the incoherent
domains increase in population for the larger values of the coupling radius. Interestingly, 
this is not the case for greater values of $\sigma$ such 
as shown in Fig.~\ref{fig-coh3}. In the latter, 
the incoherent domains gradually increase in population as the coupling ratio $d$ increases until
$d\geq 0.2$ after which point the population remains approximately constant. 
These observations  come to agreement with the imprint of the chimera 
states on the map in Fig.~\ref{fig-map}.

\par Overall, the measures of
coherence indicate that as the coupling range increases (below all-to-all coupling)
the emergence of chimera states is favored
and the mean phase velocity deviations between coherent and incoherent regions become 
more prominent. This holds for all values of $\sigma$ outside the transition region.

\subsection{Combined connectivity with variation of the coupling strength}
\label{sec:sigma}

In this subsection we systematically vary the coupling strength and discuss the effect of this tuning
on the synchronization patterns and on the chimera morphology. As before, the neurons are set
on a regular ring topology and with the same connectivity scheme, initial conditions and parameters  
as in subsection \ref{sec:radius}.

\begin{figure}[h!]
\includegraphics[clip,width=1.1\linewidth,height=1.1\linewidth,angle=0]{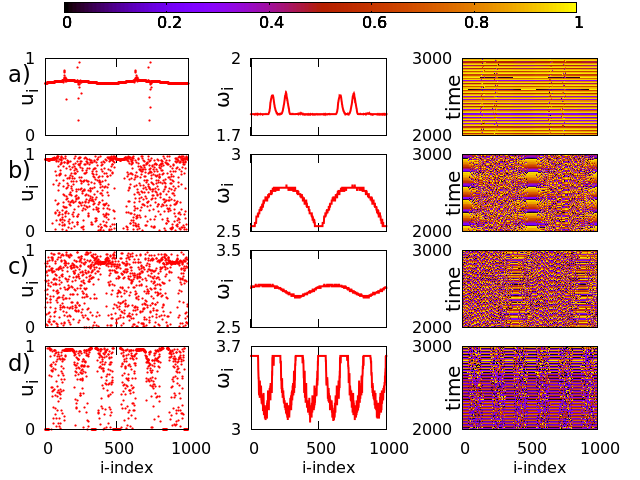}
\caption{\label{fig-s1} (Color online)
LIF system with superposition of nonlocal  and diagonal connectivity:
Typical snapshots (left column), mean-phase velocity (middle column) and space-time plots (right
column). a) $\sigma =0.4$, b) $\sigma=0.6$, c) $\sigma=0.8$ and d) $\sigma=1.6$.
Other parameters are  $R=120\quad (d=0.481)$, $N = 1000$, $\mu = 1$
and $u_{\rm th} = 0.98$. All realizations start from the same initial
conditions, randomly chosen between 0 and $u_{\rm th}$. 
}
\end{figure}

\par
We scan the system over different values of the coupling strength $\sigma$ and for coupling range $R=120$
and present the spatiotemporal evolution in Fig.~\ref{fig-s1}. 
 We first note that the mean phase velocity
$\omega$ overall increases with $\sigma$ and the solitaries developed from small $\sigma$ (Fig.~\ref{fig-s1}a)
grow gradually to form typical chimeras (Fig.~\ref{fig-s1}b).
We observe that for small values of the 
coupling strength the coherent regimes have higher $\omega$ values than the incoherent regimes (see 
Fig.~\ref{fig-s1}a,b,
 middle panels). 
Surprisingly, this behavior is inverted when $\sigma$ takes values greater than 1. In this case
the elements belonging to the incoherent regimes have 
higher $\omega$'s than in the coherent regimes (compare middle panels in Figs.~\ref{fig-s1}a,b and \ref{fig-s1}d).

\par
For values of the coupling strength in the range $1.0 \leq \sigma<1.5$ (with $R<200, \>\> d<0.801$)
complete $\omega-$synchronization is observed, where all elements acquire common mean phase velocity, but their phases
might be asynchronous. The precise size of the $\sigma$-range
where complete $\omega-$synchronization is observed depends on the value of $R$. E.g., for $R=200$ 
complete $\omega-$synchro- nization
is observed for $1\le \sigma \le 1.5$, while for other values of $R$ the range of $\sigma$ which
supports complete $\omega$-synchronization might be different.

\begin{figure}[h!]
\includegraphics[clip,width=1.2\linewidth,angle=0]{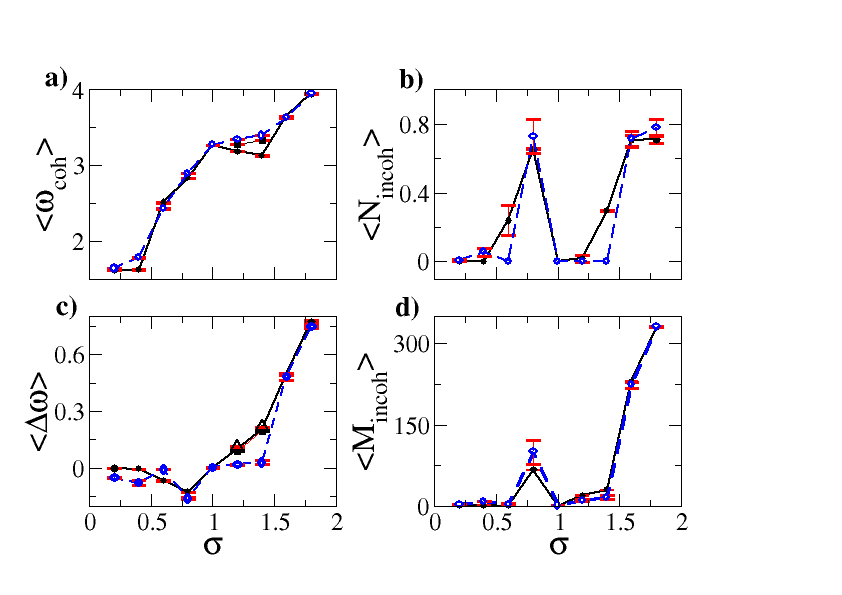}
\caption{\label{fig13} (Color online)
Measures of coherence and incoherence for different values of the coupling strength and for $R=120,\>\> d=0.481$
(blue-dashed lines) and $R=200,\>\> d=0.801$ (black-solid lines).
All other parameters are as in Fig.~\ref{fig-R}. Averages are taken over 14 initial conditions.
}
\end{figure}

Looking further to the measures of coherence and incoherence as a function of the coupling strength, Fig.~\ref{fig13},
for both $d=0.481,0.801$ we notice the increase of $<w_{\rm coh}>$ as the coupling strength becomes stronger.
For $\sigma=1.0$,  $\sigma=1.2$  and $\sigma=1.4$ the $\omega$-values do not change drastically
for finite system sizes. In these cases,
complete $\omega$-synchronization is observed followed by phase desynchronization.
Despite all elements  being completely desynchronized (they all have different phases) they do not have
significantly distinguishable $\omega$'s. Therefore, the state of complete phase desynchronization around $\sigma =1$
is characterized by coherence in respect to the distribution of the $\omega$'s.
\par  $<\Delta\omega>$ and $<M_{\rm incoh}>$ are also
increasing as $\sigma$ increases and receive their highest values  at the same point as $<\omega_{\rm coh}>$.
$<\Delta\omega >$ and $<M_{\rm incoh}>$  are equal or almost equal to zero for values of the coupling strength
$\sigma=1.0, \sigma=1.2 $ and $\sigma=1.4$, as a result of the approximately constant values of the $\omega 's$
in these cases. The number of the incoherent elements, $<N_{\rm incoh}>$, increases with the increase of $\sigma$ up to
$\sigma=0.6$ and does not significantly change thereafter. For $\sigma=1.0, \sigma=1.2 $ and $\sigma=1.4$
all elements are $\omega$-synchronous, therefore $<N_{\rm incoh}>=<M_{\rm incoh}> =0$ for these areas,
even though they are phase asynchronous.
\par
For  larger values of the coupling range, e.g.
 $R\geq 200$,
and within the same $\sigma$-range ($1.0 \leq \sigma <1.5$)
we note that the synchronized regimes are divided into two subgroups
with different frequencies. These two subgroups are shown in Fig.~\ref{fig13}a,c, where the two coherent domains
are depicted with a black circle and a black triangle for $\sigma =1.2$ and $\sigma =1.4$, respectively. 
This effect will be discussed in detail in sec.~\ref{sec:critical}.
Another observation is that in the case of ($\sigma = 1.8, R = 200$) a four-headed
 chimera appears along with two smaller incoherent regions.
Interestingly this four-headed
 chimera consists of wider incoherent/coherent regimes and of additional smaller incoherent regions similar to solitary states.
This is further discussed in Appendix \ref{append}.

\par Overall,  $\omega$-inversion between coherent and incoherent regions takes place when $\sigma$ crosses
the value 1. Around $\sigma =1$ complete $\omega-$synchronization takes place, while the precise $\sigma$-range
where complete $\omega-$synchro- nization is observed depends on the value of $R$. For even larger values
of $\sigma$ the phenomenon of two-level synchronization is observed which also depends both on $\sigma$ and $R$, and
 will be further discussed in the next section.

\subsection{Two-level synchronization chimeras}
\label{sec:critical}

In this subsection we discuss in more detail the two-level synchronization
which has been recorded for relatively large values of $\sigma$ and $R$.
As working parameters we will use $\sigma=1.2-1.5 $ and $R=200$,  $N = 1000$, $\mu = 1$
and $u_{\rm th} = 0.98$.
As demonstrated in the previous sections for values of the coupling strength $\sigma<1$ the coherent domains
of the chimera states have lower mean phase velocity values compared to the incoherent ones, 
whereas for $\sigma \ge 1.6$ the opposite is observed. Therefore, certain areas such as  [$\sigma=1.2,\>\> d=0.801$]
and [$\sigma=1.4,\>\> d=0.801$] behave like border areas and the system presents two-level synchronization.

\begin{figure}[h!]
\includegraphics[clip,width=1.0\linewidth,angle=0]{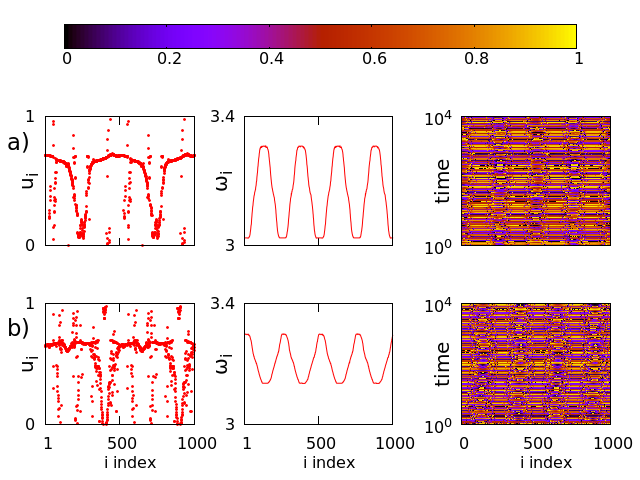}
\caption{\label{fig:14} (Color online)
LIF system with superposition of nonlocal  and diagonal connectivity:
Mean-phase velocities (left column) and spacetime plots (right
column). a) $\sigma=1.2$, b) $\sigma=1.4 $.
Other parameters are $R=200$ $(d=0.801)$, $N = 1000$, $\mu = 1$
and $u_{\rm th} = 0.98$. All realizations start from the same initial
conditions, randomly chosen between 0 and $ u_{\rm th}$.}
\end{figure}

\par Figure \ref{fig:14}  presents the mean phase velocities and spacetime plots
 for the previously mentioned parameters. Looking at the spacetime plots
in Fig.~\ref{fig:14}a (right panel) one simply
observes the formation of 8 incoherent regions mediated by 8 coherent regions. This is nothing
exceptional, 
however, by looking at the mean phase velocity profiles, one observes that the 8 coherent regions
are split in two groups: one group is characterized by high mean phase velocity, $\omega_{\rm coh1}$
while the other one is characterized with low mean phase velocity, $\omega_{\rm coh2}$. The members
of the two groups alternate, while the incoherent elements have $\omega$ 's that lay on
the sloping line between the two groups. The same observations hold for different values of $\sigma$
as shown in  Fig.~\ref{fig:14}b, but the difference between the two coherent levels vary with the
coupling constant.

\par For this two-level synchronization it is not possible to use the classical rules
described in refs. \cite{omelchenko:2015} to calculate the  measures of coherence and incoherence.
We now introduce a new algorithmic scheme which allows us to distinguish between the two regimes
of coherence: 
\begin{itemize}
\item First
we calculate the maximum $<\omega_{\rm coh1}>$ and the minimum $<\omega_{\rm coh2}>$ values of the mean
phase velocities. These
two values characterize the mean frequency in each of the two synchronization levels.
\item
For the calculations of the $N_{\rm incoh}$ we set a small tolerance (e.g., $a=0.01$), that serves
as a border between the coherent and incoherent domains.
\item We set the counter  $N_{\rm incoh}=0$ and we scan all elements, $i=1, \cdots N$, to calculate the difference
of the mean phase velocities between element $i$
and each one of the coherent domains $ <\omega_{\rm coh1}>-\omega_i$ and 
$\omega_i-< \omega_{\rm coh2}>$.
 If $ \left[ < \omega_{\rm coh1}>-\omega_i \right] > a$ 
and $\left[ \omega_i- < \omega_{\rm coh2} \right] > a$ then we consider element $i$
as belonging to the incoherent domain and
we increase the counter  $N_{\rm incoh}= N_{\rm incoh}+1$. 
\item When all elements are scanned we normalize the counter $N_{\rm incoh} \to N_{\rm incoh}/N$.
\end{itemize}

\begin{figure}[h!]
\includegraphics[clip,width=1.1\linewidth,height=1.0\linewidth,angle=0]{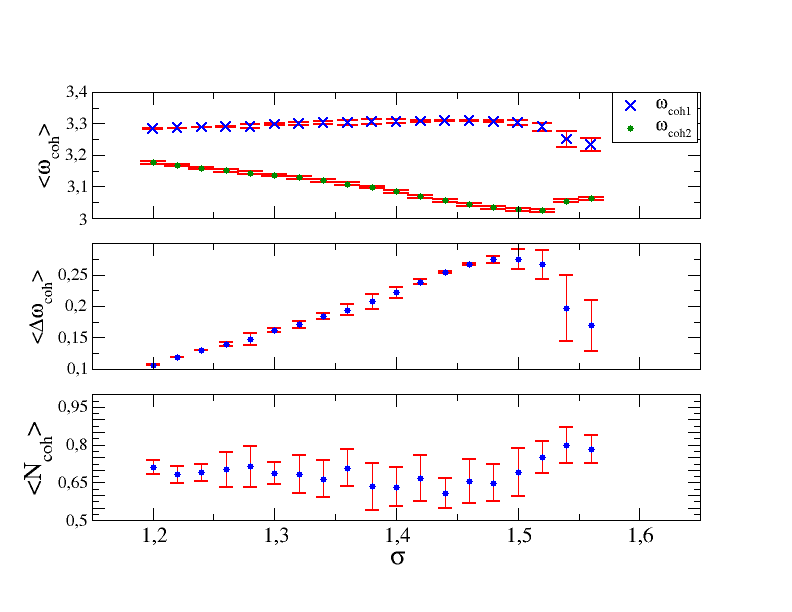}
\caption{\label{fig:15} (Color online)
Measures of coherence for LIF system with superposition of nonlocal  and diagonal connectivity 
for two-level synchronization. Parameter values are:
 $R=200$ $(d=0.801)$, $N = 1000$, $\mu = 1$
and $u_{\rm th} = 0.98$. All realizations start from initial potentials
randomly chosen between 0 and $ u_{\rm th}$ and averages are taken over 14 initial conditions.}
\end{figure}

Results of the two synchronization levels  $< \omega_{\rm coh1} >$,  $ < \omega_{\rm coh1} >$, 
 $\Delta\omega_{\rm coh} = < \omega_{\rm coh1} > - < \omega_{\rm coh2}>  $ and $ < N_{\rm incoh}> $ are
demonstrated in Fig.~\ref{fig:15} (top).
We note that the synchronization highest level,  $< \omega_{\rm coh1} > $,
does not significantly increase with the coupling strength $\sigma$ for values $1.2<\sigma <1.55$
while it decreases thereafter, see Fig.~\ref{fig:15}a. On the contrary, the lowest
synchronization  level, $< \omega_{\rm coh2}> $, decreases as $\sigma$ increases. In this interval, $1.2<\sigma <1.55$,
the difference between the $\omega $'s of the two synchronization levels increases with $\sigma$,
as seen in Fig.~\ref{fig:15} (middle row) and by
comparison between Figs.~\ref{fig:14}a and b.
The measure $<N_{\rm incoh}>$ does not show a significant change in this parameter area.
For values of $\sigma >1.55$ the two synchronization levels tend to approach one another. The latter along with the previous
observations on the increase of $<\omega_{\rm coh2}>$, suggests that the elements following this synchronization
scenario merge with the incoherent domains as $\sigma$ increases. This gradually leads to the chimera states
that appear for higher values of $\sigma$, i.e. $\sigma=1.6$, where the coherent domains have higher $\omega$
values and only one synchronization level is observed.

Based on the appearance of two sets of coherent regions with different mean phase velocities
 we propose here a possible scenario for the creation of chimera states.
 It is possible that due to the influence
of the neighbors and for some parameter regions
the coupled elements acquire bistable equilibrium states, i.e., they can be found in two
frequency regimes, called $\omega_{\rm coh1}$ and $\omega_{\rm coh2}$. Under the dynamics and due to the initial conditions
some of them tend to the equilibrium frequency $\omega_{\rm coh1}$ forming domains around them,
 while others tend to the second equilibrium $\omega_{\rm coh2}$. The elements in-between are influenced by both
domains and acquire progressive frequency values, bridging the gap. Therefore, they behave incoherently because
they have different frequencies. This scenario of bistable coupled elements can be 
used as a generic scenario for the formation of chimera
states. Even in the case where only one synchronization level is evident while the incoherent regions are 
 viewed as arc-shaped in the mean phase velocity profiles, one may imagine the presence of a second,
non-visible (not well formed) level of synchronization at the top of the arc. This scenario may be supported
by findings in ref. \cite{tsigkri:2017} where even three levels of synchronization are realized, see
sec.~4 and figures therein.

\section{Conclusions}
\label{conclusions}

In the current study we discuss  the influence of a) diagonal connectivity and b) the combination of nonlocal-diagonal
connectivity in the dynamics and especially in the appearance of chimera states  in a network of LIF
elements. We demonstrate that different connectivities influence the dynamics of the system
and especially the presence and form of chimera states.

\par We show that the multiplicity of the chimera states is modulated 
by the values of the coupling range and the coupling strength and we identify areas in the $(d, \sigma )$ parameter space
where the system develops different behaviors. We show numerically that  
 the mean phase velocity is not significantly affected by the variation of the coupling radius; 
instead, the variation of the coupling strength tunes the $\omega$ values. More specifically, the 
$\omega$'s increase overall
 as the coupling strength increases. This is intuitively expected since a high value of the coupling 
strength amplifies the intensity of the interaction between different elements, which synchronize by dragging
each other.

\par We stress the development of two new chimera patterns which emerge for relatively large values of the 
coupling constant: 
\begin{enumerate}
\item The two-level chimera states: these are states which consist of two groups of coherent states, one with
low mean phase velocity and one with high. The elements of the two groups alternate, while the incoherent
regions ensure the continuity of the $\omega$-profiles 
and serve to bridge the gap between the regions of low and high
mean phase velocities.
\item The instability which develops in the middle of the coherent regions giving rise to ``rebels'' or solitary
states. These solitary states build up to become incoherent regions as the coupling constant $\sigma$ decreases.
\end{enumerate}

\par Based on the 1st case (bi-leveled coherent domains), we propose a generic scenario 
on the formation of chimera states based on bistable element interactions.

In previous studies the influence of the refractory period in the nonlocal and reflecting connectivity cases was
studied and it was shown that it affects drastically the chimera multiplicity and synchronization patterns in general. 
Since  the refractory period is observed in biological neurons, it would be interesting to investigate its effects
in the network with diagonal and combined connectivity. 
Another open question worth investigating is whether the novel effects 
(two-level synchronization, inversion of coherence, solitary to chimera growth) caused by the combined,
nonlocal-diagonal connectivity in the LIF model are generic in many neuron models, or they are restricted to the 
LIF model which is characterized by the abrupt resettings of the potential. Combined connectivity effects are
worth studying in the FitzHugh-Nagumo oscillator,
the Hindmarsh-Rose model and Stuart-Landau oscillator and other models assimilating the activity of neurons.
Problems related to memory effects in chains of nonlinear oscillators are also relevant for further studies, in 
the sense that the final states depend highly on the initial conditions (potentials) of the oscillators. This
is a hard problem (see \cite{wu:2018-2} and references therein) when a large number of oscillators are considered
and is more tractable in small networks composed by a few oscillators.

\vskip 1.0cm

\textbf{Acknowledgements}

The authors would like to thank Prof. N. Sarlis and Dr. J. Hizanidis for helpful discussions.
 This work was supported
by computational time granted from the Greek Research
\& Technology Network (GRNET) in the National HPC
facility - ARIS - under project ID PA002002. 
NTD acknowledges financial support from the IKY Greek State Scholarship Foundation. 
\vskip 0.50cm
\textbf{Author Contribution Statements}

NTD, IK and GK performed the numerical studies. NTD and AP
participated to the designing of the study and to the coordination of the project. 
All authors contributed to the draft of the manuscript. All authors read
and approved the final manuscript.

\appendix
\section{Instabilities due to large values of the coupling constant}
\label{append}
\par
\begin{figure}
\includegraphics[width=1.05\linewidth,angle=0]{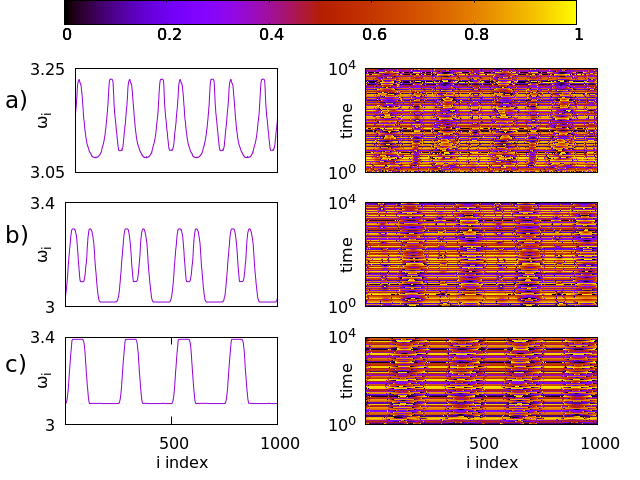}
\caption{\label{fig:17} (Color online)
LIF system with superposition of nonlocal  and diagonal connectivity:
mean-phase velocity (left)  and space-time plots (right). Parameters are:  a) $\sigma=1.3$, $R=230$, $(d=0.921)$; 
b) $\sigma=1.6$, $R=230$, $(d=0.921)$; and c)
$\sigma=1.7$, $R=230$, $(d=0.921)$.
Other parameters are $N = 1000$, $\mu = 1$,
and $u_{\rm th} = 0.98$.
}
\end{figure}

 The dynamics of the LIF model allows for increasing the coupling strength arbitrarily high. Due to the
condition \ref{eq1b} the potentials $u_i$ never diverge as they are always reset to zero when they increase
above the threshold $u_{\rm th}$. This is a particularity of the LIF system, while most oscillators become unstable
when the coupling strength $\sigma$ increases beyond a critical value. Therefore, we have the opportunity to explore
the network dynamics under very large values of the coupling strength. 
An example of the complexity induced is
an instability within the synchronous regions which leads to the formation of asynchronous regions of smaller
width than the synchronous ones. This instability is visible for large $\sigma$ values as
was also discussed in sec.~\ref{sec:critical}. The formation of the instability is visible in Fig.~\ref{fig:17}. 
Here the coupling strength values are $\sigma =1.3-1.8$, while all
other parameters are as in Fig.~\ref{fig:14}. The spacetime plots clearly indicate the 
formation of two large and two smaller incoherent regions separated by coherent regions.

 By looking at the spacetime plots, as $\sigma$ decreases from  1.7 (Fig.~\ref{fig:17}c) to 1.6 (Fig.~\ref{fig:17}b)
one instability is created in
the middle of the synchronous regions and four secondary, smaller synchronous regions appear. The size of the secondary
regions increases as $\sigma$ decreases from  1.6 (Fig.~\ref{fig:17}b) to 1.3 (Fig.~\ref{fig:17}a) and for even
smaller $\sigma$ eight coherent regions of equal sizes appear (not shown). The same effect can be also
seen in Figs.~\ref{fig:d3} and \ref{fig:d5} for different parameter values where one new asynchronous
region is created in the middle of the single synchronous domain.
 \par By looking at the
$\omega$ profiles in Figs.~\ref{fig:17} 
we note that the instabilities set in the synchronous regions having the largest mean phase
velocity $\omega_{\rm coh1}$, while the coherent regions with low mean phase
velocity $\omega_{\rm coh2}$ are not affected. The $\omega$'s in the new incoherent regions decrease gradually 
with $\sigma$ and they approach $\omega_{\rm coh2}$ when eight coherent regions of equal sizes are formed, around 
$\sigma =1.2$ (not shown). 
The mechanism causing this second instability in large values of the coupling constant is not completely
understood.

\end{document}